\documentclass[12pt]{iopart}

\usepackage{graphicx}
\usepackage{url}

\begin{document}

\title{A microkelvin cryogen-free experimental platform with integrated noise thermometry}

\author{G Batey$^1$, A Casey$^2$, M N Cuthbert$^1$, A J Matthews$^1$, J Saunders$^2$, A Shibahara$^2$}

\address{$^1$Oxford Instruments Omicron NanoScience, Tubney Woods, Abingdon,
Oxfordshire, OX13 5QX, UK}
\ead{Anthony.Matthews@oxinst.com}

\address{$^2$Department of Physics, Royal Holloway, University of London, Egham, Surrey, TW20 0EX, UK}
\ead{A.Casey@rhul.ac.uk}

\begin{abstract}
We report experimental demonstration of the feasibility of reaching temperatures below 1~mK using cryogen-free technology. Our prototype system comprises an adiabatic nuclear demagnetisation stage, based on hyperfine-enhanced nuclear magnetic cooling, integrated with a commercial cryogen-free dilution refrigerator and 8~T superconducting magnet. Thermometry was provided by a current-sensing noise thermometer. The minimum temperature achieved at the experimental platform was 600~$\mu$K. The platform remained below 1~mK for over 24~hours, indicating a total residual heat-leak into the experimental stage of 5~nW. We discuss straightforward improvements to the design of the current prototype that are expected to lead to enhanced performance. This opens the way to widening the accessibility of temperatures in the microkelvin regime, of potential importance in the application of strongly correlated electron states in nanodevices to quantum computing.
\end{abstract}

\pacs{07.20.Mc}

\submitto{\NJP}

\maketitle

\section{\label{intro}Introduction}
Quantum effects in condensed matter physics can be revealed when systems are cooled to a temperature below which the thermal energy is less than their characteristic energy scale. The long tradition of cooling condensed matter systems to progressively lower temperatures has uncovered a wealth of new phenomena. However most of the emphasis on cooling systems to below 1~mK has focussed on quantum fluids and solid. The widely agreed imperative now is to open the microkelvin range to experiments on nanoelectronic devices and strongly correlated electron systems. Examples are: fractional quantum Hall effect with potential applications in topological quantum computing \cite{Sarma2005,Marcus2008}; new strongly correlated quantum states in semiconductor nanodevices \cite{Hanson2007}; charge pumps \cite{Pepper2010}; possible nuclear spin ordering in two dimensional quantum wells \cite{Loss2007}; brute force cooling of nanomechanical resonators into the quantum regime \cite{Schwab2004}. In addition the decoherence times of superconducting qubits may ultimately depend upon the intrinsic coherence of Josephson junctions \cite{Paik2011} which should be improved at lower temperatures \cite{Catelani2011}.

In order to cool these systems into the microkelvin range, the requirements are: an appropriate low-temperature platform and thermometry; appropriate sample thermalisation. This paper considers the first of these.

Continuous cooling to the low mK regime is usually achieved using a dilution refrigerator. Whilst very powerful dilution refrigerators can operate at temperatures down to $\sim$~2~mK \cite{Vermeulen1987, Cousins1999}, and the cold liquid in dilution refrigerators can be coupled to a condensed matter experiment \cite{Samkharadze2011}, the dramatic decrease in availability (and associated increase in
cost) of $^3$He in recent years \cite{Kouzes2009} has made these very large systems uneconomical for all but the most specialised low-temperature facilities.

On the other hand, the temperature regime down to $\sim$~10~mK has recently been opened up by the availability of cryogen-free systems. Here a dilution refrigerator with a relatively small charge of helium mixture is operated with a pulse tube refrigerator pre-cooling stage \cite{uhlig2002}. These ``dry'' cryogen-free systems can be installed without the need for associated complex research infrastructure, such as a helium liquefaction plant, or in remote locations. Such systems can be automated to a higher degree than their ``wet'' counterparts, and are simple to operate. In conventional systems, with liquid helium precooling, the overall diameter is constrained by the neck of the helium dewar, which influences the rate of liquid helium ``boil-off''. In cryogen-free systems this restriction does not apply. It allows experimental platforms typically several hundred mm in diameter. This has enabled a range of more complex services to be installed onto such refrigerators; for example bulky signal-conditioning elements such as cryogenic amplifiers, microwave components (bias-tees, circulators, switches etc.) and filtering (such as metal powder filters) which have proved invaluable for experiments aimed at quantum information processing.

Nuclear cooling \cite{Andres1982} offers the possibility of extending experimental temperatures into the $\mu$K regime, well below those accessible to even the most powerful dilution refrigerators. A variety of strategies have been developed \cite{Pickett1988, Pobell2007}. The most popular is the adiabatic nuclear demagnetization of copper, precooled in a (large) magnetic field (required to generate the initial entropy reduction in the nuclear refrigerant) by a thermal link connected to a dilution refrigerator with an unloaded base temperature of \textless~10~mK \emph{via} a superconducting heat switch. In this case the demagnetisation cooling process is``single-shot'', as only a finite amount of energy can be absorbed by the nuclear stage as the system warms.
These restrictions have not proved to be a problem in practice as the duty cycle between the pre-cooling and demagnetisation stages of a typical experimental run are short compared to the low-temperature hold time, provided the heat leak into the system is sufficiently small.
Recently, semiconductor nano-structures have been cooled using an array of nuclear refrigerators attached to
measurement leads \cite{Clark2010}. However, it is not obvious that this approach could be implemented for a wide range of condensed matter physics experiments, especially those requiring complicated wiring arrangements.

The approach adopted here is the integration of cryogen-free dilution refrigerators and superconducting
magnets \cite{Batey2009}, with the entire system running from a single pulse tube cooler, with a ``bolt-on'' nuclear refrigerator to extend the accessible temperature range to below 1~mK, thus maintaining compatibility with a wide range of experimental applications. Given the extreme precautions taken with nuclear adiabatic demagnetization cryostats to create ultra-low mechanical noise environments in order to minimize heat leaks due to vibration of the nuclear stage, the feasibility of realising this goal was not apparent; this has been addressed in the work reported here. 

\section{\label{fridge}The cryogen-free system}
The pulse tube coolers used on cryogen-free dilution refrigerators are known to be a source of mechanical noise, particularly on systems where the room temperature, first ($\sim$~50~K) and second stage ($\sim$~3~K) components of the pulse tube coldhead (the volume in which the gas expands) are not vibrationally decoupled from the dilution refrigerator \cite{pelliccione2013}. Initial measurements of cryogen-free refrigerator and magnet systems \cite{Batey2009} did not show a significant increase in the base temperature of the refrigerator when the installed magnet was persistent at its full field of 12~T. However in those tests the mixing chamber plate only experienced the fringing field of the magnet, not the full field to which any nuclear refrigerant would be exposed. To our knowledge it has not been demonstrated previously that this potential obstacle to the implementation of nuclear demagnetisation techniques in a cryogen-free environment could be overcome.

The choice of refrigerant for nuclear cooling has been reviewed \cite{Pickett1988, Pobell2007}. Here we consider two options: copper or PrNi$_5$. Copper is recognised as being the refrigerant of choice for the attainment of the lowest possible temperature, but in order to generate a significant entropy reduction in the refrigerant the starting conditions are demanding: a 9~\% entropy reduction can be achieved by cooling to a temperature of $\sim$~10~mK in a magnetic field of $\sim$~8~T. The high bulk electrical conductivity of copper at low temperatures coupled with the large magnetic field and the possibility of mechanical vibrations in a cryogen-free system led to copper being rejected as the refrigerant in this work. A large entropy reduction can be achieved in PrNi$_5$ at higher temperatures and in lower applied fields due to the hyperfine-enhancement of the field experienced by the nuclei; in a field of 6~T a 70~\% entropy reduction is reached at 25~mK. Additionally, the poor bulk conductivity of the material (comparable to that of brass) should minimise eddy current heating resulting from any motion of the refrigerant in the field. The nuclear spin ordering in PrNi$_5$ also allows the possibility to demagnetize to zero magnetic field; subsequent warming of the stage is governed by the Schottky heat capacity maximum which occurs at $\sim$~0.5~mK \cite{Kubota1980}, this contrasts with the use of copper, where nuclear ordering is only observed at the much lower temperature of 58~nK \cite{Huiku1982}, and so in the temperature regime of the present application the system is a nuclear paramagnet and the nuclear heat capacity is proportional to the square of the final magnetic field.

In view of the concerns about the possible effects of vibrational heating, both in the pre-cool phase and the warm-up following demagnetization, we opted for PrNi$_5$ as our coolant. The dilution refrigerator used in this work, an Oxford Instruments Triton~200, has been described previously \cite{Batey2009}. The typical performance parameters are a cooling power in excess of 200~$\mu$W at 100~mK, and a base temperature below 10~mK. We used a standard Triton~200, with the only anti-vibration measures being the decoupling of the pulse tube cooler first and second stages from the refrigerator plates using flexible copper braids.

\begin{figure}
\includegraphics[width=0.95\textwidth]{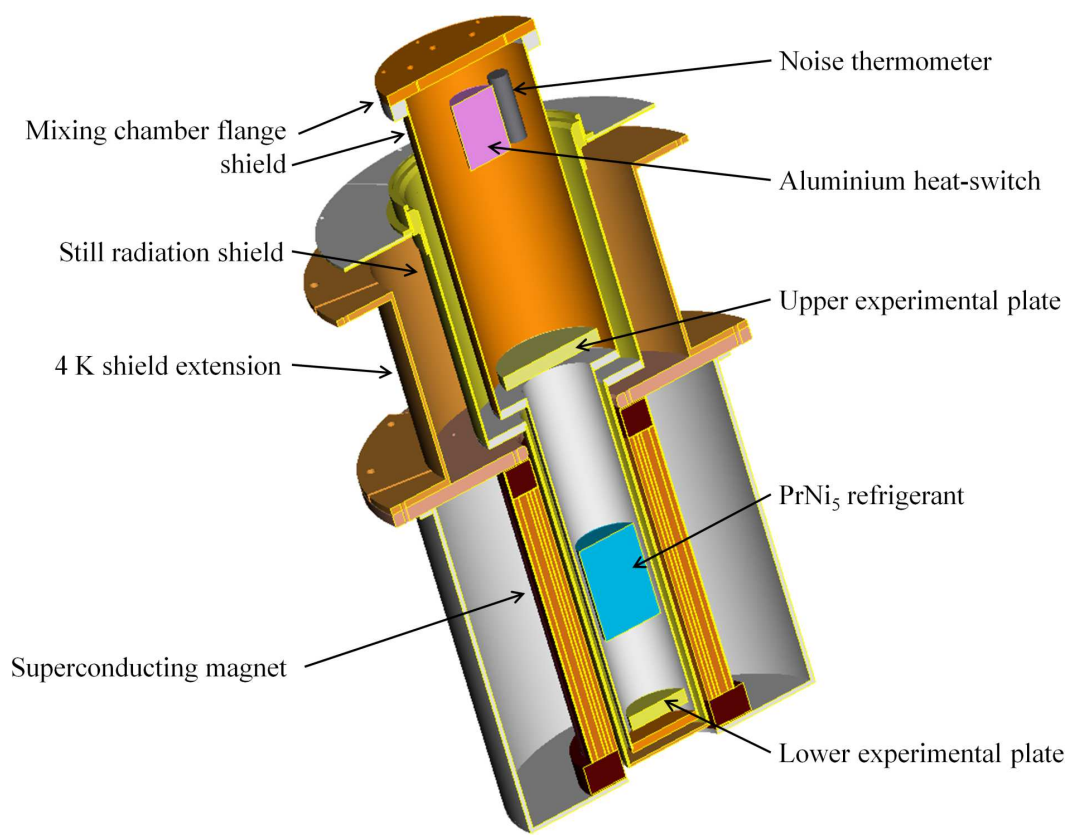}
\caption{\label{fig:fridge} A cross-section view of the layout of the demagnetisation stage below the mixing chamber. The 4~K shield extension allows the superconducting magnet to be mounted in a lower position with respect to the mixing chamber. The superconducting heat switch and noise thermometer are positioned just below the mixing chamber flange. The low thermal conductivity support structure of the demagnetisation stage is not shown for clarity and nor are the thermal links, which are described in the main text.}
\end{figure}

The layout of the demagnetisation stage below the mixing chamber is shown in figure~\ref{fig:fridge}. The superconducting magnet available for this work was of standard solenoid design, with a 77~mm cold bore and a maximum field of 8~T when operated at 5~K. The field compensation of this solenoid was not optimised; however the stray field at the mixing chamber and heat switch was lowered by the simple expedient of increasing the distance between the mixing chamber plate and the centre of the solenoid. This increase in cryostat length was also necessary to accommodate the nuclear refrigeration stage. The overall system height was \textless~2.5~m, compatible with standard laboratory space.

The nuclear stage consists of 128~g of the inter-metallic compound PrNi$_5$ in the form of nine 6~mm diameter $\times$ 50~mm long rods, which were pre-tinned with 99.99~\% cadmium. Each rod was connected to the upper and lower experimental plates (figure~\ref{fig:fridge}) with 1~mm diameter copper wire with a residual resistance ratio $\sim$~1000. One wire per rod extends to the upper plate and eight wires per rod to the lower plate. The stage used in this work was adapted from a pre-existing stage made at Cornell \cite{Parpia1985}. The upper experimental plate was connected to the mixing chamber through a superconducting aluminium heat switch \cite{Lawson1982}. The switch had six 1~mm diameter silver wires bonded to either end, which in turn were diffusion bonded to copper blocks which made mechanical connections to the upper experimental plate and the mixing chamber. This link was
particularly long, $\sim$~30 cm, to mitigate the effects of the non-optimal stray field, however it was still possible to precool the stage to below 20~mK in a 6.2~T field within 24~hours. The heat load on the mixing chamber during the precool was a combination of eddy current heating of the nuclear stage and components of the mixing chamber exposed to the large stray fields of the magnet, in particular the brass mixing chamber radiation shield which experienced the full field of the magnet. With the heat switch open, it was possible to cool the mixing chamber to $\sim$~15~mK.

\section{\label{thermo}The current-sensing noise thermometer}

Attaining temperatures significantly below dilution refrigerator temperatures poses an experimental challenge. Measuring the temperatures so-attained poses another. The base temperature of dilution refrigerators can be established with nuclear orientation thermometry \cite{Marshak1983}, however this slow technique is not suitable as a practical thermometer for regular use, and is not usable at $\mu$K temperatures.
Other convenient methods of temperature measurement, such as carbon resistance sensors \cite{Samkharadze2010}, shot noise in tunnel junctions \cite{Spietz2006} or Coulomb blockade \cite{Pekola1994} also prove difficult to implement below a few mK. The most common approach is to rely on the Curie-law paramagnetism of nuclear spins \cite{buchal1978} or the magnetism of dilute electronic paramagnets \cite{Paulson1979} calibrated with a fixed point device. Other methods are possible if a sample of superfluid $^3$He is part of the set-up \cite{Todoschenko2002}.

Here we have deployed a current sensing noise thermometer \cite{Casey2003}. This thermometer exploits the low intrinsic noise and high sensitivity of a dc Superconducting Quantum Interference Device, SQUID, configured as a current amplifier to detect the Johnson noise produced by a resistive element. The resistive sensor is mounted on a copper platform linked to the demagnetisation stage. The sensor is connected to the SQUID input inductance \emph{via} a superconducting twisted pair. The mean square noise current flowing in the SQUID input coil per unit bandwidth, arising from the thermal noise in the resistor, is then given by
\begin{equation} \label{eq:noise}
\left\langle I^{2}_{N} \right\rangle = \frac{4k_{B}T}{R}\left(\frac{1}{1+\omega^{2}\tau^{2}}\right)
\end{equation}
where $\omega = 2 \pi f$ and the time constant $\tau = L/R$. Here $L$ is dominated by the input coil inductance of the SQUID $L_{i}$; the resistance $R$ can be chosen either to optimise speed of measurement or reduce the noise temperature of the measurement system. In this work a sensor resistance of 0.24~m$\Omega$ was chosen, when coupled to an extremely sensitive 2-stage SQUID \cite{Drung2007} this results in a current noise power due to the SQUID of $2.5\times10^{-25}$~A$^{2}$/Hz or equivalently an amplifier with a noise temperature of $T_{N}=1$~$\mu$K. For comparison we note that a thermometer prepared in a similar way mounted on a traditional copper nuclear demagnetisation cryostat has been cooled to below 200~$\mu$K in our laboratory. We also remark that noise thermometers optimised for operation in the dilution refrigerator temperature range can achieve 1~\% precision in 100~ms of measurement time.

\begin{figure}
\centering {\includegraphics[width=0.65\textwidth]{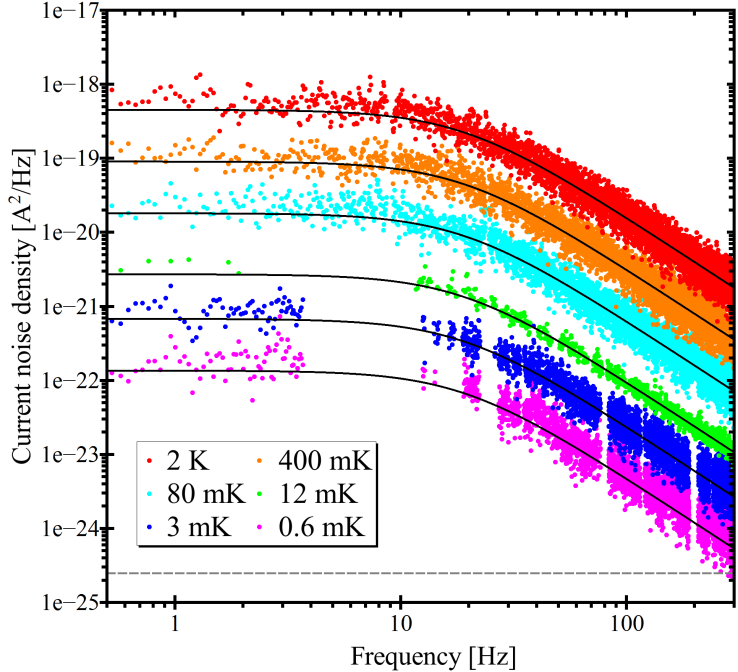}}
\caption{\label{fig:thermom} Current noise spectra from the 0.24~m$\Omega$ noise thermometer spanning four orders of magnitude of temperature from 2~K down to sub~mK. Each trace consists of 10 averages of 20~s traces, taken at 50~kS/s, giving a total acquisition time of 200~s per trace. The 12~mK trace (green) was 40 averages of 5~s traces. Solid black lines are fits to the data using the form given in equation \ref{eq:noise}. The dashed grey line is the intrinsic SQUID noise floor extrapolated from high frequencies.}
\end{figure}

Typical examples of the current noise measured over a range of temperatures are shown in figure~\ref{fig:thermom}. For these measurements we have mounted the noise thermometer in a region of low stray field near to the heat switch; see figure~\ref{fig:fridge}. It is thermally connected to the lower nuclear stage plate through a link consisting of 37 $\times$ 0.7 mm diameter, annealed copper wires. Extraneous noise peaks arising from environmental noise have been filtered. Scatter in the measured frequency-dependent noise power is higher than that observed on a conventional (non cryogen-free) refrigerator, which may be a quieter environment due to the absence of microphonic and / or triboelectric noise arising from vibrations caused by the pulse tube cooler. Nevertheless we obtain a precision of 2~\% in a measurement time of 200~s from fits to the power spectrum.

\section{\label{results}Results}
\begin{figure}
\centering {\includegraphics[width=0.65\textwidth]{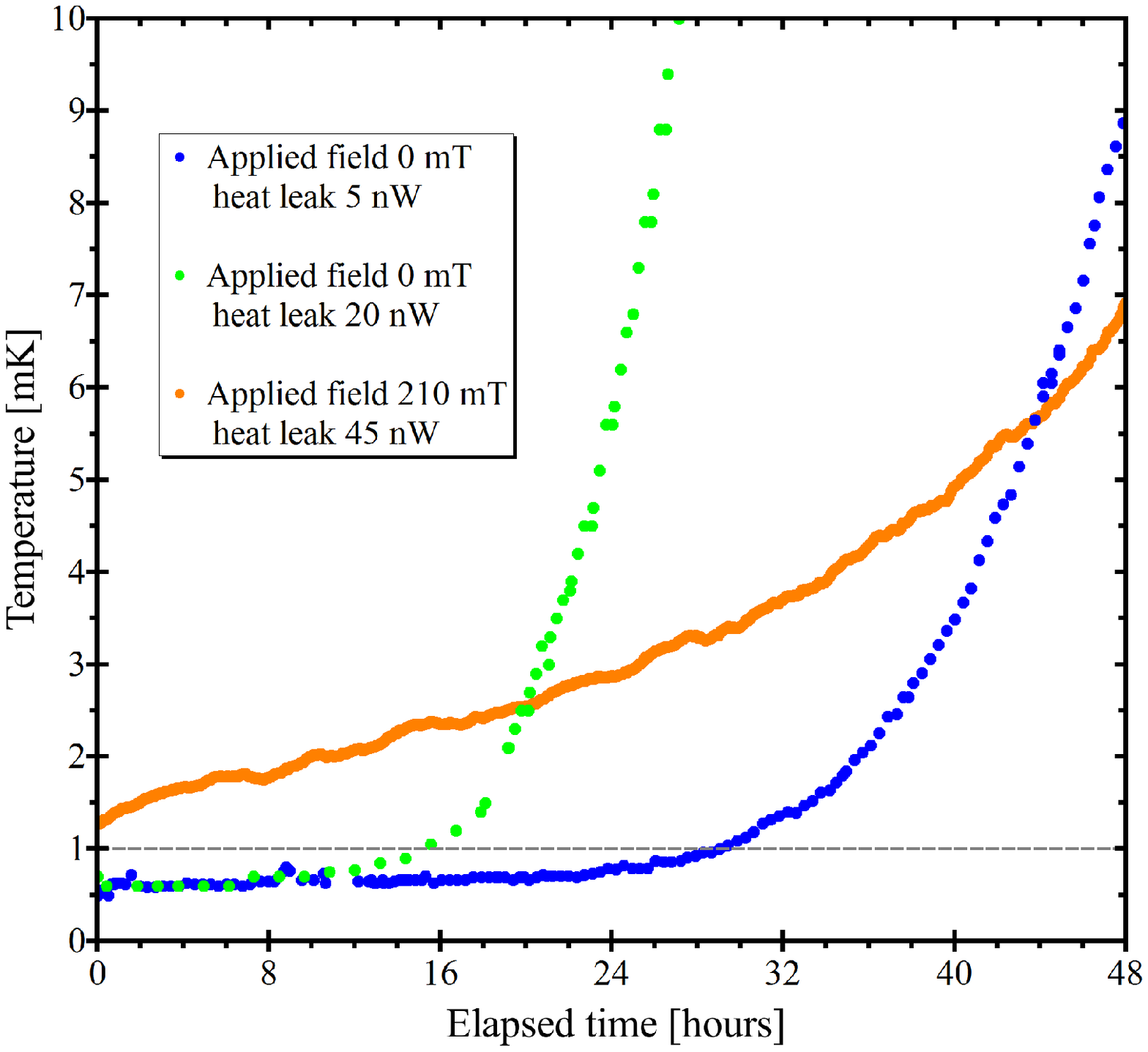}}
\caption{\label{fig:results1} Temperature versus time following three warm-up curves after demagnetising the stage. On one run a heat leak of 20~nW was present, demagnetising to zero applied field resulted in the warm-up shown by the green circles, here the thermometer was cooled to a temperature of 0.6~mK. The subsequent demagnetisation was stopped in an applied field of 210~mT (orange) demonstrating the possibility to trade final temperature versus hold time, here the heat leak was $\sim$~45~nW in this field. The lowest heat leak achieved was 5~nW; under this heat leak a warm-up in zero applied field (blue) had a hold time below 1~mK $\sim$~30 hours.}
\end{figure}

Here we present the results from this prototype system, which demonstrate the feasibility of the approach, despite a number of non-optimal conditions, discussed in the conclusion. The precool field was limited to \textless~6.2~T by the tolerable upper limit to the stray field at the superconducting heat switch. Typically a 24~hour precool was sufficient to cool the stage to 20~mK, achieving an entropy reduction of 80~\% of the free spin value \cite{Folle1981}. The superconducting heat switch was then opened and the demagnetisation carried out in a series of steps. The field was halved at each step and the rate was adjusted so that each step took the same amount of time, until a final step from 100~mT to zero applied field. Hence demagnetising to zero field could be achieved in around 6~hours.

\begin{figure}
\centering {\includegraphics[width=0.65\textwidth]{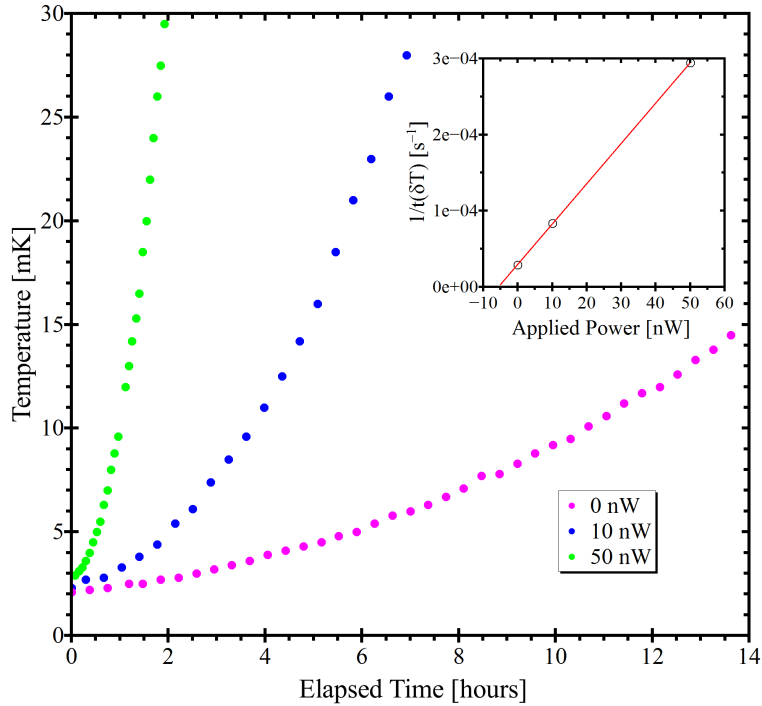}}
\caption{\label{fig:results2} The intrinsic heat leak to the demagnetisation stage in zero applied field can be determined by observing the time, t, required to warm the system for a given change in temperature, $\delta$T, as a function of applied power. The inset shows 1/t($\delta$T) for warming from 2 to 12~mK. The intrinsic heat leak of 5~nW is given by the intercept of a linear regression to these data.}
\end{figure}

Figure \ref{fig:results1} shows the results of such a demagnetisation on two separate cryostat cool-downs. In one case an intrinsic heat leak to the stage of 20~nW was observed. The heat leak was determined by applying additional electrical power to the stage and monitoring the time required to warm it over the temperature range from 2 to 12~mK, figure \ref{fig:results2}. We found the heat leak was sensitive to the precise arrangement of the flexible lines around the pulse tube refrigerator cold-head and to the addition of lead shot damping masses ($\sim$~10~kg) to the top of the cold-head itself. Whilst the intrinsic heat leak could be reduced in this way no correlation could be observed with the vibrational spectra measured with accelerometers \cite{HSJ}, sensitive in the frequency range 1-1000~Hz, installed on the refrigerator 3~K stage. The background heat leak was improved to 5~nW, comparable to that achieved in many conventional copper nuclear demagnetisation systems. Heat leaks below the nW level can only be achieved by taking the utmost care in vibration isolation and choice of materials used in construction \cite{Buck1990}.
In the run where a 20~nW heat leak was observed the thermometer was cooled to 600~$\mu$K. After reaching this temperature it stayed below 1~mK for 16~hours. An increased hold time can be achieved at the expense of a higher minimum temperature by stopping the demagnetization at a higher final field. In figure \ref{fig:results1} we show the results of a demagnetisation to 210~mT with a lowest temperature of 1.2~mK. In this case the stage remained below the base temperature of the dilution refrigerator for over two days. Finally, with an improved heat leak of 5~nW, a hold-time below 1~mK of over 24~hours was achieved.

\section{\label{conc}Conclusions}
We have presented initial results obtained on a cryogen-free dilution refrigerator with a bolt-on PrNi$_5$ nuclear demagnetisation stage. The results: a base temperature of 600~$\mu$K and a residual heat leak 5~nW, indicate that cryogen-free systems are indeed suitable environments for microkelvin experiments despite the concerns over the vibrational stability of such systems. Pre-cooling the nuclear refrigerant to temperatures \textless~20~mK in a field of $\sim$~6.2 T is accomplished in around 24~hours, with the demagnetisation taking a further 6~hours. The hold time below 1~mK can be \textgreater~24 hours. This represents a reasonable duty cycle for many experiments in this regime.

Such a simple demagnetization stage could, in principle, be added to any cryogen-free dilution refrigerator system that is suitable for the operation of a superconducting magnet. The temperatures were determined using a current-sensing dc SQUID noise thermometer that allowed us to measure directly sub-mK temperatures.

The present set-up is subject to a number of relatively straightforward modifications that are anticipated to lead to improvements in performance. These include optimisation of magnet design to provide appropriate field cancellation regions. In this work the nuclear stage was added to an existing cryogen-free dilution refrigerator for which there was no particular attention devoted to vibration isolation. Remotely mounting the turbo-molecular pump, used for the $^3$He circulation, and decoupling the pulse tube cooler from the system top-plate will reduce the levels of vibrational noise, together with improvements in the rigidity of the refrigerator mounting frame.

Cryogen-free sub-mK platforms do therefore seem be a realistic prospect, dramatically improving the accessibility of the ultra-low temperature frontier for exploration and discovery. The addition of a research magnet, to enable the combination of a high magnetic fields and microkelvin temperatures remains a future technical challenge to be addressed.

\section*{References}
\bibliographystyle{unsrt}
\bibliography{1mK}

\ack{The authors would like to thank Prof. Jeevak Parpia for supplying the PrNi$_5$ stage used in this work, J. Elford who contributed to the design and procurement of prototype shields and other parts necessary for these initial demagnetisation tests, and H. van der Vliet for noise thermometer tests. The superconducting aluminium heat switch was kindly supplied by R. Haley, Lancaster Cryogenics Ltd. The development of the noise thermometer was supported by the EU through the FP7 Capacities Specific Programme, MICROKELVIN project number 228464 and the European Metrology Research Program (EMRP). The EMRP is jointly funded by the EMRP participating countries within EURAMET and the European Union.}

\end{document}